\documentclass[manuscript]{emulateapj}
\usepackage{textcomp}
\usepackage{amssymb}
\usepackage[normalem]{ulem}
\usepackage{leftidx}
\usepackage{amsmath}
\usepackage{graphicx}
\shorttitle{A Universal SFDM halo mass: Andromeda and MW dwarfs}
\shortauthors{Lora V.}

\begin{document}
\title{A Universal SFDM halo mass for the Andromeda and Milky Way's dSphs?}
\author{V. Lora\altaffilmark{1}}
\affil{Astronomisches Rechen-Institut, Zentrum f\"{u}r Astronomie der Universit\"{a}t Heidelberg, \\
             M\"{o}nchhofstr. 12-14, 69120 Heidelberg, Germany}

\email{vlora@ari.uni-heidelberg.de}

\begin{abstract}
Dwarf spheroidal galaxies are the most common type of galaxies, and 
are the most dark matter dominated objects in the Universe. Therefore, 
they are ideal laboratories to test any dark matter model. 
The Bose-Einstein condensate/scalar field dark matter model
considers that the dark matter is composed by spinless-ultra-light 
particles which can be described by a scalar field. This model is an 
alternative to the $\Lambda$-cold dark matter model.
In this work I study the kinematics of the dwarf spheroidal satellite 
galaxies of the Milky Way and Andromeda, under the scalar field/BEC dark 
matter paradigm in two limits: when the self interacting parameter is 
equal to zero, and when the self interacting parameter is $\gg1$. 
I find that dwarf spheroidal galaxies with very high mass-to-light ratios 
(higher than $100$) are in better agreement with an NFW mass density profile.
On the other hand, dwarf spheroidal galaxies with relatively low 
mass-to-light ratios and high luminosities are better described with the 
SFDM model. Such results are very encouraging to further test alternative 
dark matter models using the dynamics of dwarf galaxies as a tool.
\end{abstract}

\section{Introduction}

Dwarf spheroidal (dSph) galaxies are believed to be the most common type of galaxies 
in the Universe, and the building blocks of more massive galaxies in hierarchical 
formation scenarios.The large mass-to-light ratios of these dSph galaxies suggest 
that the dynamics of such galaxies are dominated by dark matter (DM). Therefore, 
the study of dSphs is crucial for a better understanding of the nature of DM. In particular, 
the dSph of the Local Group (LG) of galaxies satellites of the Milky Way (MW) and 
Andromeda galaxies, are good targets for study, since individual stars can be resolved and 
evolutionary histories can be derived in great detail.

\cite{walker:09} proposed that all dSphs follow a universal mass profile,
which means that all dSphs are embedded within a universal dark matter halo. 
They found the best-fit for the relationship between velocity 
dispersion and half-light radius for a maximum velocity $V_{max} = 13$~km/s, 
and a scale radius, $r_0 = 150$~pc, for a cored DM halo. For the NFW profile, the best 
fit is achieved for $V_{max} = 15$~km/s, and a scale radius $r_0 = 795$~pc.

More recently, \cite{collins:14} extended the previous work, incorporating kinematic information 
of $25$ Andromeda dSph \citep{collins:13}. They found a discrepancy between the DM 
density profile best-fit for the dSphs in the MW compared with that for Andromeda's dSphs (for
both, cored and NFW cases). A good agreement between both MW dSphs and Andromeda 
dSphs samples is only reached, when three Andromeda outliers (And XIX, And XXI and 
And XXV) and three MW outliers (Hercules, CVnI and Sagittarius) are removed from the 
dwarf galaxy sample.

The discrepancy between the cuspy density profiles of DM halos predicted in simulations, 
with the cored density profiles derived from observations of dSph galaxies and Low 
Surface Brightness galaxies \citep{bosch:00,kleyna:03,blok:02,lora:09,walker:11,amorisco:12,
jardel:12} (the cusp/core problem), the overpopulation of dark substructure \citep{klypin99}, 
and the to-big-to-fail problem \citep{read:06,boylan:11,boylan:12}, have motivated alternative
DM candidates to the $\Lambda$CDM model.

An alternative model that lately has gained interest, is to consider that the DM is 
made of bosons described by a real (or complex) scalar field $\Phi$: the Bose-Einstein condensate/scalar 
field DM model (BEC/SFDM) \citep{sin:94, ji:94, jaeweon:96, peebles:99, matos:00, 
guzman:00,matos:09,magana:12b}. The BEC/SFDM model is also known as Fuzzy DM \citep{hu:00} or 
recently as Wave DM \citep{schive:2014}. 

A very interesting feature of the BEC/SFDM model, is that it naturally 
produces cored halos. The size of the cores of such halos (for a fixed 
self-interacting parameter  $\Lambda$) depends on the mass of the BEC/SFDM boson 
and the mass of the SFDM halo ($M_{DM}$).

The SFDM model has been proved to be very successful \citep{magana:12a}. It is consistent with the 
anisotropies of the cosmic microwave background radiation (CMB) \citep{rodriguez:10},
and it can also achieve a better fit to high-resolution rotation curves of
low-surface-brightness galaxies \citep{robles:12}, compared to the  NFW \citep{nfw} 
profile.

Lately, \cite{lora:12} and \cite{lora:14} used the internal stellar structures of 
dSph galaxies to establish a preferred range for the mass $m_{\phi}$ of the 
bosonic particle. They performed $N$-body simulations and explored how the 
dissolution time-scale of the cold stellar clump in Ursa Minor (UMi) and Sextans depends on 
$m_{\phi}$. They found that for a boson mass in the range of $(3<m_{\phi}<8)\times10^{-22}$~eV, 
the BEC/SFDM model would have large enough cores to explain the stellar substructure
in dSph galaxies.

Moreover, \cite{diez-tejedor:14} find a preferred scale radius of $\sim0.5-1$~kpc, 
from the kinematics of the eight brightest dSphs satellites of the MW. 

In this work I investigate the idea of a universal mass profile for the dSph 
population of the MW \citep{walker:09} and Andromeda (M31) under the SFDM model, 
and compare it with a cored and NFW profile \citep{collins:14}.  

The article is organized as follows: in \S \ref{sec:SFDM} the SFDM model is described 
and  the Schr\"{o}dinger-Poisson system is briefly reviewed.
In \S\ref{sec:dwarfs}, I discuss the universal NFW and cored 
DM profile for the dSphs in the MW and M31. In \S\ref{sec:results}, I describe 
our results, and finally, in section \S\ref{sec:conclusions} I discuss the results and 
give our conclusions.


\section{The BEC/SFDM halos}
\label{sec:SFDM}
The DM halos can be interpreted as BEC/SFDM Newtonian gravitational configurations 
in equilibrium, which can be described by the so called Schr\"odinger-Poisson system, 
which is the Newtonian limit of the Einstein-Klein-Gordon equations 

\begin{equation}
 i\hbar\frac{\partial\psi}{\partial t} = -\frac{\hbar^2}{2m_{\phi}} \nabla^2 \psi + 
U m_{\phi} \psi    
+ \frac{\lambda}{2m_{\phi}} |\psi|^2\psi \mbox{ , } 
\label{schroedingerA}
\end{equation}

\begin{equation}
 \nabla^2 U = 4\pi G m_{\phi}^{2} \psi \psi^\ast  \mbox{ . } 
\label{poissonA}
\end{equation}

The critical mass for such configurations is 
$M_{crit}\sim 0.6 \left( \frac{ m_{P}}{m_{\phi}}\right)\sim10^{12}$~M$_{\odot}$ 
\citep{ruffini:69,seidel:91}.

In Equations~\ref{schroedingerA} and \ref{poissonA}, $m_{\phi}$ is the mass of the boson 
associated with the wave function $\psi$. $U$ is the gravitational potential (produced by the 
mass density source $\rho=m_{\phi}^2|\psi|^2$), and $\lambda$ is the self-interacting parameter. It has been 
shown that the latter term determines the compactness of the structure \citep{guzman:06}. 

It is convenient to work with the dimensionless Schr\"odinger-Poisson system, given as follows
\citep{ThesisArgelia}
\begin{equation}
 i\frac{\partial \hat{\psi}}{\partial \hat{t}} = -\frac{1}{2} \hat{\nabla^2} \hat{\psi} + 
\hat{U} \hat{\psi} + \hat{\Lambda} |\hat{\psi}|^2 \hat{\psi} \mbox{ , } 
\label{schroedinger_nounits}
\end{equation}

\begin{equation}
 \hat{\nabla}^2 \hat{U} =  \hat{\psi} \hat{\psi}^\ast  \mbox{ ,} 
\label{poisson_nounits}
\end{equation}
where
\begin{equation}
\label{lambda}
\Lambda=\frac{\lambda}{4 \pi m_{\phi}^{2} G }
\end{equation}
 
The solutions to the Schr\"odinger-Poisson system relevant for this work
are those with spherical equilibrium \citep{ThesisKevin,ThesisArgelia}. Hence, it is
convenient to work with the Schr\"odinger-Poisson system in spherical coordinates

\begin{equation}
\label{schrodinger_spherical}
  i\frac{\partial \hat{\psi}}{\partial \hat{t}} = -\frac{1}{2\hat{r}} 
\frac {\partial^{2}}{\partial \hat{r}} (\hat{r} \hat{\psi}) + 
\hat{U} \hat{\psi} + \hat{\Lambda} |\hat{\psi}|^2 \hat{\psi} \mbox{ , } 
\end{equation}

\begin{equation}
\label{poisson_spherical}
 \frac {\partial^{2}} { \partial \hat{r}} (\hat{r}\hat{U}) = \hat{r} \hat{\psi} \hat{\psi}^\ast  \mbox{ .} 
\end{equation}

The solutions of the Schr\"odinger-Poisson system are obtained assuming an harmonic 
behavior of the scalar field, such that
\begin{equation}
\hat{\psi}(r,t)=e^{-(i \hat{\gamma} \hat{t})}\hat{\phi}(\hat{r}) \mbox{  ,} 
\end{equation}
where $\hat{\gamma}$ is an adimensional frequency.

Substituting $\hat{\psi}$ in Equations~\ref{schrodinger_spherical} and \ref{poisson_spherical},
the Schr\"odinger-Poisson system now reads as

\begin{equation}
 \frac{d^2}{d\hat{r}^{2}}(\hat{r}\hat{\phi})= 2 \hat{r} (\hat{U}-\hat{\gamma}) 
+ 2 \hat{r} \Lambda \hat{\phi}{^3} \mbox{ ,}
\label{S-icA} 
\end{equation}
\begin{equation}
 \frac{d^2}{d\hat{r}^{2}}(\hat{r}\hat{U})= \hat{r}\hat{\phi}^2 \mbox{ .}
\label{P-icA}
\end{equation}

The SFDM halos are constructed by obtaining ground state (stable) solutions of the 
Schr\"odinger-Poisson system (\ref{S-icA}-\ref{P-icA}) \citep{guzman:04,ThesisArgelia,lora:12,lora:14}. 
To guarantee regular solutions, the boundary conditions must satisfy that at $r=0$
$\partial_{r} U=0$, $\partial_{r} \phi=0$, and that $\phi(0)=\phi_c=$, where $\phi_c$
is an arbitrary value.

The mass of this BEC/SFDM halo can be estimated as
\begin{equation}
M= 4\pi \int^{\infty}_{0} {{\phi}}^2 {r}^2 d {r} \mbox{ ,}
\label{eq:masa_r}
\end{equation}
and must be finite number. The radius of this configuration is defined as
$r_{95}$, which is the radius containing $95\%$ of the mass. Note that both 
properties, the mass and the radius of the SF halo depend on the boson 
mass, and the self-interacting term. 
Three parameters $\phi_c$, $m_{\phi}$ and $\Lambda$, define a model completely.

\begin{figure*}
\begin{center}
\includegraphics[width=.55\linewidth]{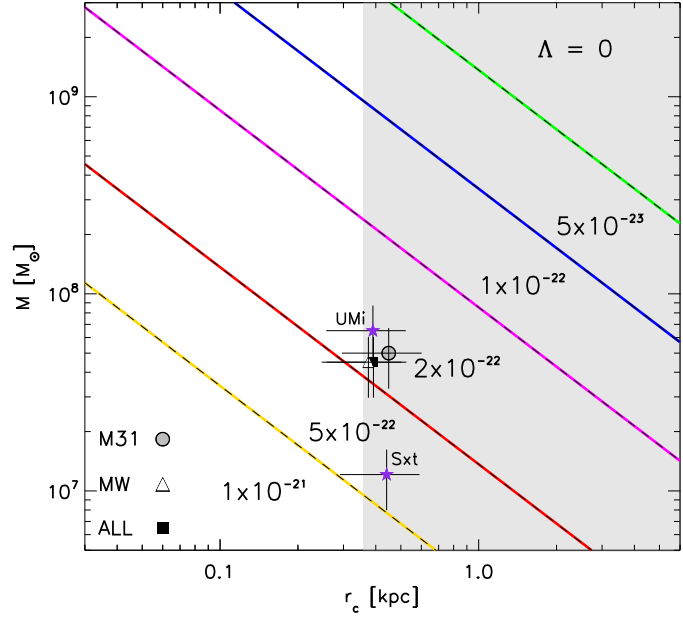}
\caption{Mass of the SFDM halo, as a function of the SFDM core radius for the $\Lambda=0$ case. 
Each diagonal line corresponds to a different value of the SFDM boson $m_{\phi}$: $5\times10^{-23}$(green), 
$10^{-22}$(blue), $2\times10^{-22}$(pink), $5\times10^{-22}$(red), and $10^{-21}$(yellow)~eV.
The shaded region shows the permitted core radius $r_{c}$, such that the stellar clumps in Ursa Minor 
and Sextans are not destroyed 
\citep{lora:12,lora:14}.}
\label{fig:FIG1}
\end{center}
\end{figure*}

\subsection{The $\Lambda\gg1$ case}
\label{sec:big_lambda}
When the self interaction between the bosons in a Bose-Einstein condensate is taken into
account, the Schr\"odinger equation can be interpreted as the mean-field approximation at 
zero temperature of the Gross-Pitaevskii equation (see Equation ~\ref{schroedingerA}).
The limit where the number of particles in the BEC is very large, and thus the self 
interacting term dominates, is called the Thomas-Fermi limit (TFL) \citep{pitaevskii:61,dalfovo:99}.

As the number of particles in the BEC becomes infinite, the TFL approximation becomes 
exact \citep{barcelo:05}, giving as a result the classical limit of the theory.
Then, the equations describing the static BEC in a gravitational field with potential
$U$ take the following form \citep{boehmer:07}:

\begin{equation}
   \nabla P\left(\frac{\rho}{m_{\phi}}\right)=-\rho \nabla \left(\frac{U}{m}\right) \mbox{ ,}\\
\end{equation}
\begin{equation}
   \nabla^{2} U =4 \pi G \rho \mbox{ .}
\end{equation}

With appropriate boundary conditions, one can integrate the latter equations, assuming an
equation of state of the form $P=P(\rho)$.
Assuming the non-linearity in the Gross-Pitaevskii equation
of the form $g(\rho)=\alpha \rho^{\Gamma}$ (where $\alpha$ and $\Gamma$ are constants greater
than zero) \citep{boehmer:07}, then the BEC equation of state is given by 

\begin{equation}
 P=P(\rho)=\alpha(\Gamma -1) \rho^{\Gamma} \mbox{ .}
\end{equation}
 
If $\Gamma$ is represented by $\Gamma=1+\frac{1}{n}$ (where $n$ is the polytropic index),
then the equation of state of the static gravitationally bounded BEC is described by 
the Lane-Emden equation

\begin{equation}
 \frac{1}{\xi^{2}} \frac{d}{d\xi} \left(\xi^{2} \frac{d\theta}{d\xi}\right) + \theta^{n}=0 \mbox{ .}
\end{equation}

In particular, for a static BEC in the TFL with a politropic index $n=1$, the Gross-Pitaevskii
equation reduces to a Lane-Emden equation of the form

\begin{equation}
 \frac{d^2\theta}{d\xi^{2}}+\frac{2}{\xi} \frac{d\theta}{d\xi}+\theta =0 \mbox{ ,}
\end{equation}
with an analytical solution
\begin{equation}
  \theta(\xi) =  \frac{sin \xi}{\xi} \mbox{ .}
\end{equation}

Using the solution given by \cite{boehmer:07} for the density profile, one has

\begin{equation}
\rho(r)= \left\{
\begin{array}{ll}
 \rho_{0} \frac{sin (\pi r /R_{max})}{\pi r /R_{max}}  & \mbox{if } r \leq R_{max} \\
 0 & \mbox{if } r > R_{max} \mbox{ ,}
\end{array}
\right.
\label{eq:TFL}
\end{equation}
\textbf{where $R_{max}$ can be interpreted as the size of the core radius of the SFDM halo.}

Finally, from Equation~\ref{eq:masa_r} one can obtain the mass within a radius $r$ 
in the TFL
\begin{equation}
 M(r)=\frac{4}{\pi}\frac{\rho_{0} R_{max}^{3}}{r} \left[ sin\left(\frac{\pi r}{R_{max}} \right)  - 
 \frac{\pi r}{R_{max}} cos\left(\frac{\pi r}{R_{max}} \right)\right] \mbox{ .}
\end{equation}

Thus, the quantities that describe a system completely in the $\Lambda\gg1$ limit (TFL)
are the central mass density $\rho(c)$ and the size of the SFDM halo $R_{max}$. 

As \cite{diez-tejedor:14} point out, the size of the DM halos in the TFL can be expressed as
\begin{equation}
 R_{max}=48.93 \left( \frac{\lambda^{1/4}}{m_{\phi}} \right)^{2}
\end{equation}
showing the dependency of $R_{max}$ with $\left(\frac{m_{\phi}}{\lambda^{1/4}}\right)$.

\section{NFW and cored profiles for the  Milky Way and Andromeda dSphs}
\label{sec:dwarfs}

The measurements of central velocity dispersions of dSph galaxies very well constrain
the dynamical mass within the deprojected half-light radii ($M_{1/2}$) \citep{walker:09,wolf:10},
then one can compare such dSph mass ($M_{1/2}$ Vs $r_{1/2}$), with a certain integrated DM mass 
profile, given a total DM halo mass $M$.   
 
\cite{walker:09} applied the Jeans equation to the velocity dispersion profiles of eight 
of the brightest dSph galaxies of the MW and consider the hypothesis that all dSphs follow 
a universal mass profile, i.e. that all dSphs are embedded within a universal dark matter 
halo. 

Using the density Equation,
\begin{equation}
\label{eq:core-nfw}
 \rho(r)=\frac{\rho_{0}}{(r/r_{0})^{\alpha} (1+r/r_{0})^{\beta-\alpha}} \mbox{  ,}
\end{equation}

\cite{walker:09} fit a cored ($\alpha=0$, $\beta=3$) and an NFW \citep{nfw} 
($\alpha=1$, $\beta=3$) DM profile.
For the cored DM halo, they found the best-fit for the relationship between velocity 
dispersion and the half-light radius, for a maximum velocity $V_{max} = 13$~km/s, 
and a scale radius, $R_S = 150$~pc. For the NFW profile, the best fit is achieved 
for $V_{max} = 15$~km/s, and a scale radius $R_S = 795$~pc.

\cite{collins:14} extended the idea of a universal mass profile, now including the M31 
objects into the analysis \citep{collins:13}. 

\begin{table*}
 \centering
 \caption{Parameters of the best fit after a CMA optimization to the SFDM profile for three groups: 
 the MW dSph(19), the M31 dSph(22), and ALL the sample(41).}
 \medskip
 \begin{tabular}{@{}cccccccccc@{}}
 \hline
 Group&Number         & M    & m$_{\phi}$     & $\epsilon$ & $r_{95}$  &  V$_{c}$  & $r_{core}$  \\
      &  of galaxies  &[$10^{7}$~M$_{\odot}$] & [$10^{-22}$~eV]& $10^{-5}$&  [kpc]    &  [km/s]   &   [kpc]       \\
  \hline
& &  &  & & & &  \\
M31& 22 & $5.12$ & 3.80 &7.07 & 0.90 & 16.15 & 0.46  \\
MW & 19 & $4.44$ & 4.54 &7.31 & 0.73 & 16.71 & 0.37  \\
ALL& 41 & $4.80$ & 4.17 &7.28 & 0.80 & 16.63 & 0.41  \\
    & &  &  & &  & \\ 
  \hline
    & &  &  & &  & \\ 
    & &  &  & &  & \\ 
  \end{tabular}
  \label{table:1}
 \end{table*}

\cite{collins:14} also use the NFW and core mass density profiles.
They obtain the best fit for the NFW (cored) profile, to the whole population of dSphs, of 
$V_{max}=14.7\pm0.5$~km/s ($V_{max}=14.0\pm0.4$~km/s) and a scale radius of 
$R_S=876\pm284$~pc ($R_S=242\pm124$~pc). They concluded that neither value is a good 
fit for many of the LG dSphs. They found that if they do not take into account five outliers, 
three in M31 (And~XIX, And~XXI, and And~XXV), and two in the MW (Hercules and CVn~I), the fit is 
significantly better for the NFW (core) mass profile, $V_{max}=16.2\substack{+2.6 \\ -1.7}$~km/s 
($V_{max}=15.6\substack{+1.5 \\ -1.3}$~km/s) and $R_S=664\substack{+412 \\ -232}$~pc 
($R_S=225\substack{+70 \\ -55}$~pc).

\section{A universal SFDM halo?}
\label{sec:results}
\subsection{The $\Lambda=0$ case}
One can consider the self-interacting term $\Lambda$ to be zero in Equation
\ref{schroedingerA} (i. e. the self-interaction is negligible), such a case is also 
known as the fuzzy DM model \citep{hu:00}.

In Figure~\ref{fig:FIG1}, the mass of the SFDM halo is plotted as a function of the 
core radius ($r_c$) for $\Lambda=0$. The core radius is defined as the radius at which 
the initial density has dropped by a factor of two.

\textbf{When analyzing the behavior of the different models with fixed $M_{DM}$ and 
$m_{\phi}$, I found out that there is a correlation between the mass of the SFDM halo 
and the core radius, for a given $m_{\phi}$. The mass of the SFDM halo is inversely 
proportional to the DM core radius independently of the value of $m_{\phi}$ (i. e. 
$M_{DM} \propto r_{c}^{-1}$), but the intercept varies depending on the value of 
$m_{\phi}$. Then, I found a fit over different values of $m_{\phi}$ and their corresponding
intercepts ($\alpha$), and found the relation between the intercept and the value of 
$m_{\phi}$ of $\alpha \propto m_{\phi}^{8.7}$.}

The complete relation between the mass of the dark 
matter halo and the dark matter halo's core radius for a given $m_{\phi}$, is given by 
\begin{equation}
M_{DM}=10^{\alpha} r_{c}^{-1} \mbox{ ,}
\end{equation}
where $\alpha$ depends on the mass of the boson as
\begin{equation}
 m_{\phi}=10^{-13.9} \alpha^{-8.7} \mbox{ .}
\end{equation}
For a given $M_{DM}$ and $m_{\phi}$, there is a slope $\alpha$, for which a unique 
$r_{c}$ will be defined. With the latter relations, one can build a set of continuous 
models for a given $M_{DM}$ and $m_{\phi}$. The continuous models are shown as diagonal 
colored lines in Figure~\ref{fig:FIG1}. Each color represent a mass of the boson $m_{\phi}$ 
(green $\rightarrow 5\times10^{-23}$~eV, blue $\rightarrow 1\times10^{-22}$~eV, pink 
$\rightarrow 2\times10^{-22}$~eV, red  $\rightarrow 5\times10^{-22}$~eV, yellow 
$\rightarrow 1\times10^{-21}$~eV).

It is clear to see that for a given halo mass, lower values of $m_{\phi}$'s have 
larger core-radius, and higher values of $m_{\phi}$ have smaller core-radius. For example, 
for a fixed SFDM halo mass of $10^8$~$M_{\odot}$, if  $m_{\phi}=10^{-22}$~eV the corresponding 
core radius is $\sim3.5$~kpc. But if $m_{\phi}=10^{-21}$~eV the corresponding core radius 
is $\sim30$~pc.

I consider three different groups of dSph data; the dSphs in the MW ($19$ galaxies), the 
dSph in the M31 ($22$ galaxies), and \scriptsize{ALL} \normalsize the galaxies ($41$ galaxies). 
In order to compare with \cite{collins:14} results, I do not take into account five dSphs 
(Sagittarius, Hercules, CvnI, AndXIX, AndXXI and And XXV).

In order to see how well these three groups of dSph galaxies can be fit with a single SFDM 
profile (for a SFDM halo mass $M$ and a SFDM boson mass $m_{\phi}$), I use the following maximum 
likelihood fitting routine
\begin{multline}
\label{eq:max_prob}
L_{SFDM} (\{r_{h,i},V_{c,i},\delta_{V_{c},i}\}|m_{\phi},M) = \\
\prod_{i=0}^{N} \frac{1}{\sqrt{2\pi} \delta_{V_{c},i}} 
\times exp \Biggl[-\frac{(V_{c,SFDM}-V_{c,i})^{2}}{2\delta_{V_{c},i}^2}\Biggr] ,
\end{multline}

where $V_{c,SFDM}$ is the circular velocity predicted by the SFDM model, $r_{h,i}$ is the 
half-light radius of the $i-$ dSph, $V_{c,i}$ is the measured circular velocity at the 
half-light radius, and $\delta_{V_{c},i}$ is its uncertainty (the observational data are
taken from \citeauthor{mcconnachie:12} \citeyear{mcconnachie:12}, \citeauthor{collins:14} 
\citeyear{collins:14} and \citeauthor{martin:15} \citeyear{martin:15}). 

Equation~\ref{eq:max_prob} is a measure of how well a model with specific parameters
$m_{\phi}$ and $M$ represents a given group of galaxies. In order to find the best fit 
for the three samples I use the Evolution Strategy with Covariance Matrix Adaptation 
(CMA-ES) optimization. For details on the CMA-ES routine, see Appendix~\ref{appendix}.

The best fit for the 19 MW sample is $M=5.12\times10^7$~$M_{\odot}$ and a $m_{\phi}=3.8\times10^{-22}$~eV. 
For the 22 M31 sample the best fit corresponds to $M=4.44\times10^7$~$M_{\odot}$ and 
a $m_{\phi}=4.53\times10^{-22}$~eV. For the whole sample ($41$) the best fit corresponds
to $M=4.8\times10^7$~$M_{\odot}$ and $m_{\phi}=4.17\times10^{-22}$~eV (see Table~\ref{table:1}).
The MW fit gives as a result a slightly more massive DM halo, and a smaller
mass of the boson, compared with the M31 and the \scriptsize{ALL} \normalsize data sets. As a 
consequence the core radius for the MW sample, is the largest ($r_c\sim0.45$~kpc). Such a size 
of the DM core radius is large enough to explain the longevity of the old-cold stellar clump 
in UMi \citep{lora:12}, and the two stellar substructures in Sextans \citep{lora:14}.

\begin{figure*}
\begin{center}
\includegraphics[width=.99\linewidth]{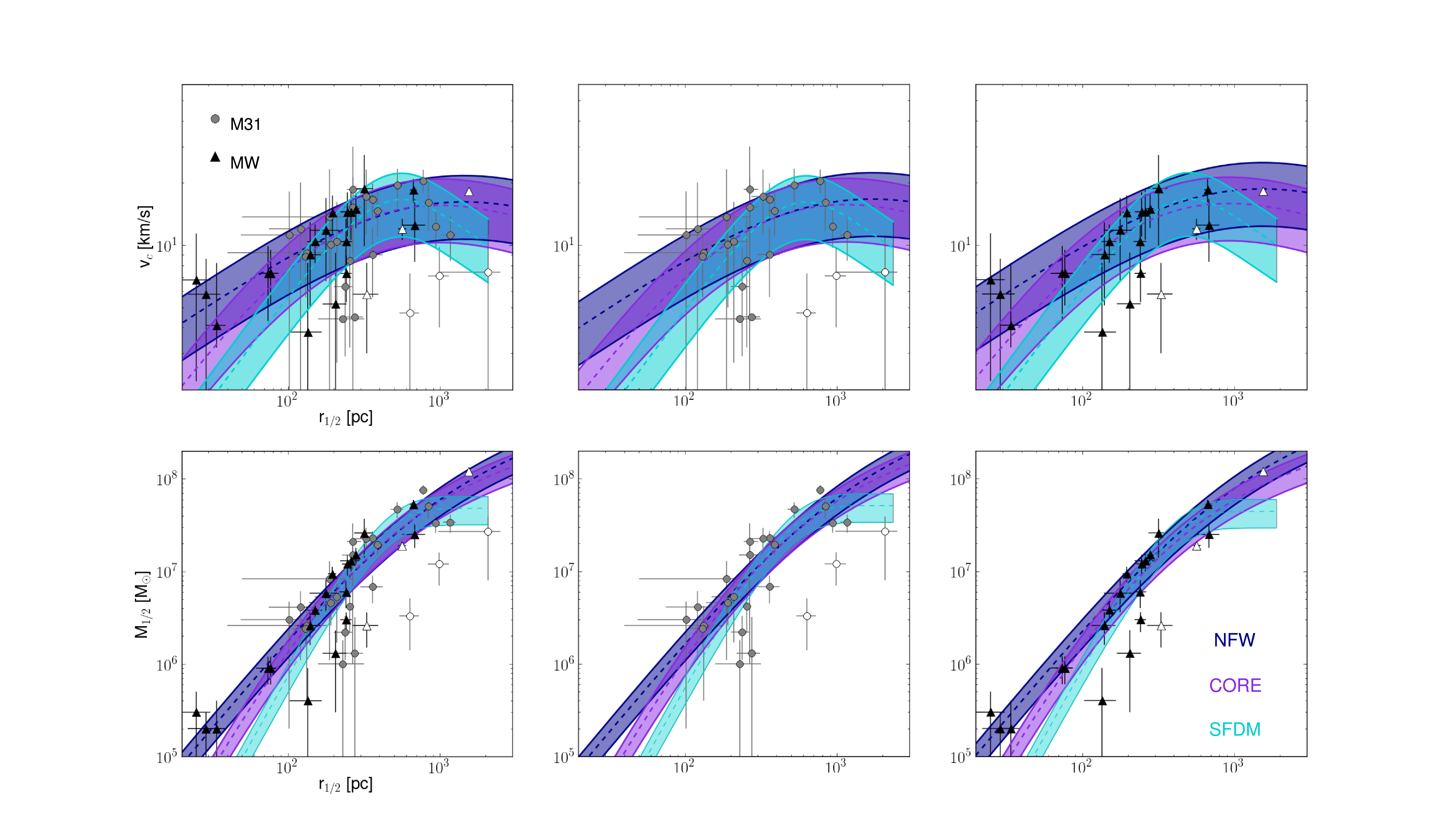}
\caption{Best fit to the NFW (dark blue), and the core (purple) DM density profile from \cite{collins:14},
leaving out the MW dSphs' outliers (Hercules, CVnI and Sagittarius, see white triangles), and three M31
 outliers M31 (AndXIX, AndXXI, AndXXV, white circles).
The SFDM profile studied in this work, is shown in light-blue. The shaded regions correspond to a
$1\sigma$ deviation, from each DM model fit. 
The upper panels show the circular velocity as a function of the half-light radius, for ALL, M31, and
MW samples, respectively.
The lower panels show the DM halo mass, as a function of the half-light radius, for ALL, M31, and
MW samples, respectively. The dSph galaxies used in this work are over plotted with their corresponding 
uncertainties. The gray circles correspond to the M31 dwarfs, and the black triangles correspond to the 
MW dSphs.}
\label{fig:FIG2}
\end{center}
\end{figure*}

In the upper panels of Figure~\ref{fig:FIG2}, I show the half-light radius circular velocity 
($V_{c,1/2}=\sqrt{\frac{GM_{1/2}}{r_{1/2}}}$) as a function of the half-light radius of all
M31 and MW dSphs in the sample, along with the best fit to \scriptsize{ALL} \normalsize sample 
for an NFW density mass profile (see dashed dark-blue line) and a cored density mass profile 
(see dashed purple lines) \citep{collins:14}, and the SFDM profile (see light-blue dash line). 
The shaded regions in Figure~\ref{fig:FIG2} indicate the $1\sigma$ error. The computed maximum 
circular velocity for the SFDM model for the \scriptsize{ALL} \normalsize sample is 
$V_{max}\approx16.63$~km/s. This value of the maximum circular velocity is in a very good 
agreement with both computed values of the NFW profile 
($V_{max}=16.2\substack{+2.6 \\ -1.7}$~km/s), and the core profile 
($V_{max}=15.6\substack{+1.5 \\ -1.3}$~km/s). 

The dSph galaxies with low $V_{c}$s and small half-light radii can be well reproduced 
with the cored and the NFW density profile, but cannot be well reproduced with the SFDM 
profile. On the other hand, the dSphs with $V_{c}$s between $6-10$~km/s, and high values 
of the half-light radii ($\sim1000$~pc) can be reproduced with the SFDM model, but 
cannot be well reproduced with either the NFW profile, or with the core profile.

In the lower panels of Figure~\ref{fig:FIG2}, I show the same fits (core, NFW and SFDM) for 
the mass of the DM halo, as a function of the half-light radius. It is very clear from this
Figure, that the NFW fits the \scriptsize{ALL} \normalsize data better at 
small-mass-small-half-light-radius, whereas the SFDM fits better the 
\scriptsize{ALL} \normalsize data at high-mass-high-half-light-radius, where both core and 
NFW profiles fail.

\begin{figure*}
  \centering
  \begin{tabular}{cc}
    \includegraphics[width=80mm]{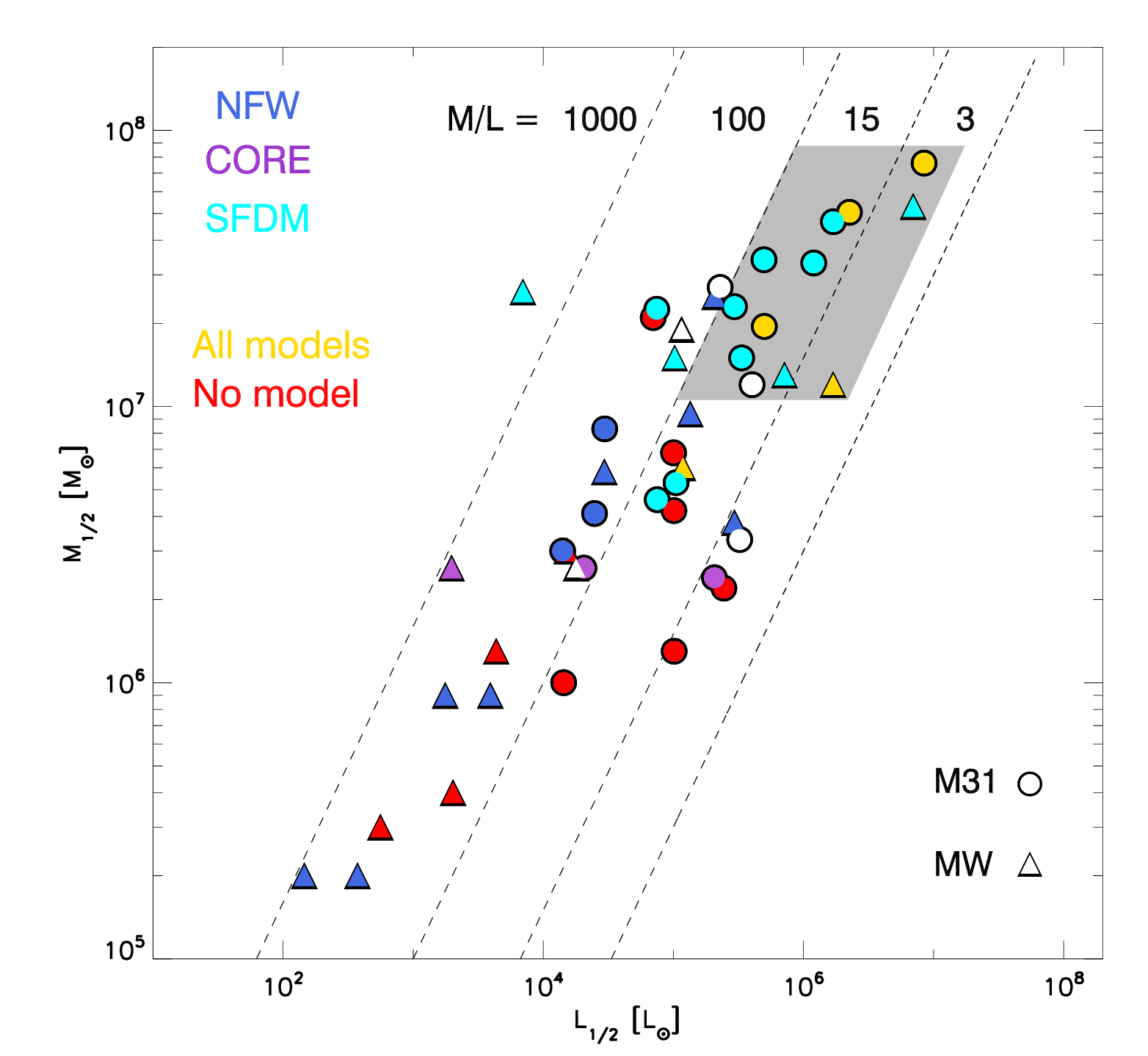}&
    \includegraphics[width=80mm]{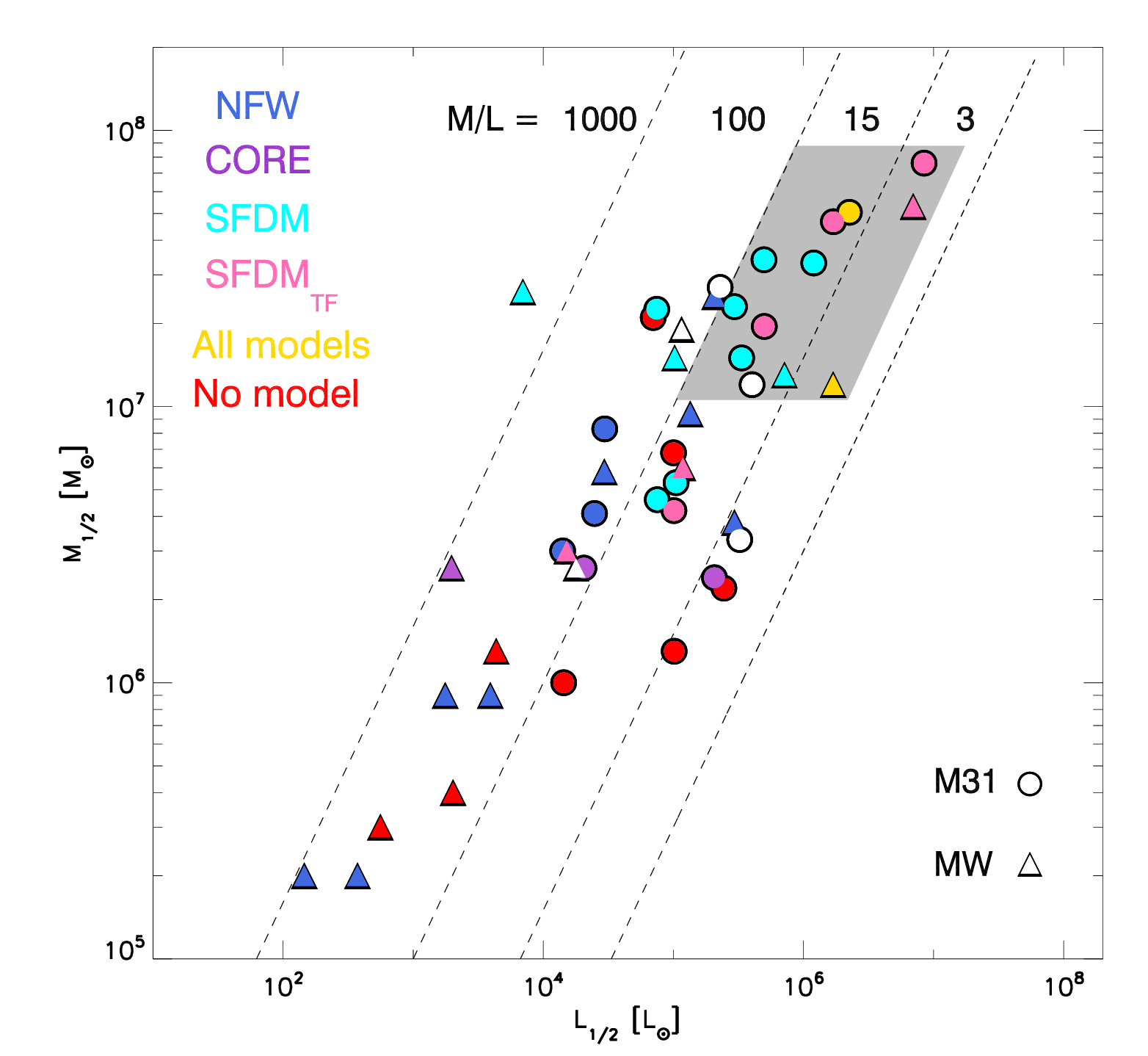}\\
  \end{tabular}
  \caption{Mass at the half-light radius as a function of the luminosity at the half-light radius. 
The M31 dSphs are shown with circles and the MW dSphs are shown with triangles. The dark-blue(purple) 
symbols show the dSphs that are in better agreement with an NFW (cored) DM profile. The light-blue symbols
show the dSphs which are in better agreement with the SFDM profile. The yellow symbols show the dwarfs
that are well reproduced with the three DM models, and the red symbols show the dwarfs that are not well
reproduced with any DM model. The white symbols represent the galaxies that were not taken into account
(Hercules, CVnI, AndXIX, AndXXI and AndXXV). The diagonal dashed lines show the mass-to-light ratios 
$M/L=1000$, $100$, $15$, and $3$.
In the right panel, the dSph that are in a better agreement
with the SFDM in the TFL i. e. $\Lambda>>1$, are also plotted (see pink symbols).}
\label{fig:FIG3}
\end{figure*}

In order to look in more detail at the properties of the dSphs that are in better agreement with 
the SFDM profile, I plot the mass inside the half-light radius as a function of the luminosity 
at the same radius (see left panel of Figure~\ref{fig:FIG3}). The diagonal dashed lines in 
Figure~\ref{fig:FIG3} correspond to four fixed mass-to-light ratios ($M/L=1000$, $100$, $15$
and $3$). The circles represent the M31 dSphs, and the triangles represent the MW dSphs.

For each dSph I computed the ratio $\zeta$ between the theoretical mass values from the 
best DM fits of the \scriptsize{ALL} \normalsize data set, and the observed mass
values \citep{collins:14}, $\zeta=M_{DM}/M_{dSph}$.  
I then define that a dSph is well reproduced by a certain DM model if $\zeta$ is at most 2.
The latter means that the dSph mass obtained from the best DM model
differs from the observed value by a maximum of a factor 2.  A dSph galaxy is in agreement
with an specific DM model when $\zeta$ is minimized.

In the left panel of Figure~\ref{fig:FIG3}, the color code is as follows: the light-blue symbols 
(both M31 circles and MW triangles) represent the dSphs which are in good agreement with the 
SFDM profile. The purple symbols show the dSphs which are better reproduced with the cored DM 
profile, and the dark-blue symbols represent the dSphs which are in  better agreement with the 
NFW profile. On the other hand, the yellow symbols correspond to the dSph that are well reproduced 
with the three DM models, whereas the red symbols show the dwarf galaxies that cannot be reproduced 
with any DM model studied in this work (i. e. $\zeta>2$).

One can immediately see from Figure~\ref{fig:FIG3} that the dSphs which are in a better 
agreement with an NFW DM profile are very dark matter dominated, with mass-to-light ratios 
between $1600\gtrsim M/L \gtrsim 100$. There are only two exceptions LeoI and LeoII, which mass-to-light 
ratios of $7.1$ and $12.9$, respectively. 

\begin{table*}
 \centering
 \caption{Parameters of the best fit in the TFL ($\Lambda>>0$) for three groups: 
 the MW dSph(19), the M31 dSph(22), and ALL the sample(41).}
 \medskip
 \begin{tabular}{@{}cccccccccc@{}}
 \hline
 Group &Number     &$\rho_{0}$             &$R_{max}$& M$_{max}$  & $\frac{m_{\phi}^4}{\lambda}$&$\frac{m_{\phi}}{\lambda^{1/4}}$&V$_{c}$ & $r_{core}$ \\
       &of galaxies&[$10^7$~M$_{\odot}$~kpc$^{3}$] &[kpc]& [$10^{7}$~M$_{\odot}$]    &[eV$^{4}$]                &[eV]                     &[km/s]  & [kpc] \\
  \hline
&  &  & & & & & & \\
M31 & 22& $8.06$ & $0.86$ & $6.716$& $3180.96$ & $7.51$ & $18.24$ & $0.52$ \\
MW  & 19& $17.96$& $0.59$ & $4.80$ & $6788.80$ & $9.08$ & $18.64$ & $0.36$ \\
ALL & 41& $9.74$ & $0.85$ & $7.60$ & $3325.20$ & $7.59$ & $19.61$ & $0.51$ \\
    & & &  &  &  & & &  \\
  \hline
    & &  &  & &  & \\ 
  \end{tabular}
  \label{table:2}
 \end{table*}

The galaxies which are better reproduced with an NFW profile are also the less 
luminous of the sample with $L \lesssim 10^{4}$~L$_{\odot}$. The only outlier is UMaI, which
is in a better agreement with the SFDM model. On the other hand, it seems that these galaxies 
have a mass at the half-light ratio limit of $\sim10^7$~M$_{\odot}$.

The upper-right gray shaded region in Figure~\ref{fig:FIG3} contains a set of five M31 dSphs,
and two MW dSphs which are in good agreement with the SFDM profile (light-blue symbols). If one
also takes into account the yellow symbols, which are dSphs well reproduced with either DM model, then 
four more can be added, resulting in a set of $11$ galaxies which are very well reproduced 
with the SFDM  model. These galaxies share the property of being less DM dominated than those
which are better reproduced with an NFW DM profile. The mass-luminosity ratios of these galaxies 
range from $\sim100$ to $5$. On the other hand, these dSphs are the most luminous galaxies of the 
sample, with a luminosity ranging from $\sim10^{5}$ to $5\times10^{6}$~L$_{\odot}$.
It is remarkable that there are no dSph in this region which are more in agreement with an NFW profile.

\begin{figure*}
\epsscale{1.2}
\plotone{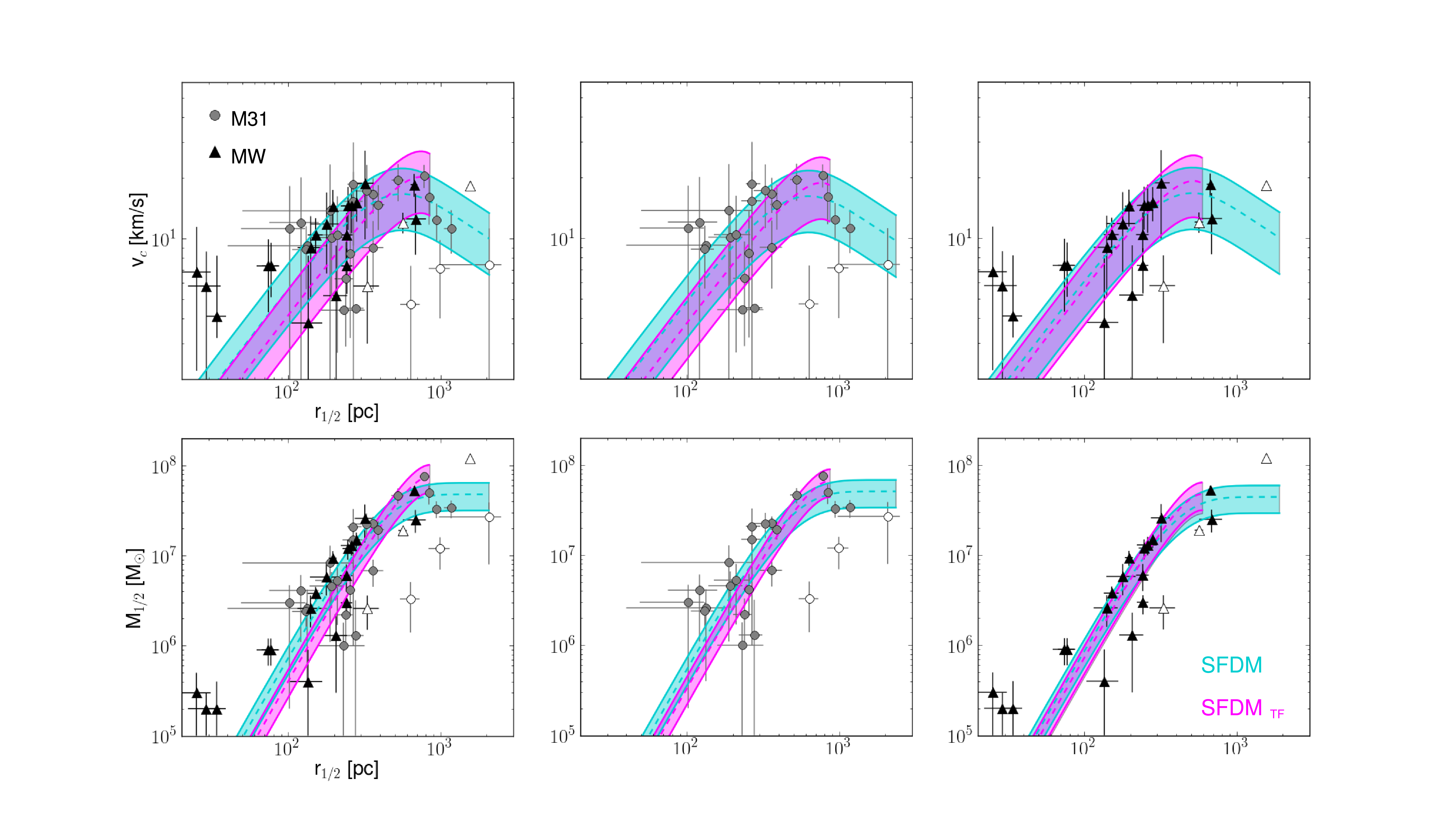}
\caption{Same as Figure~\ref{fig:FIG2} but for the best fits of the SFDM model, and the SFDM
in the TFL.}
\label{fig:FIG4}
\end{figure*}

\subsection{The $\Lambda>>1$ case}
\label{sec:Big lambda}
In this subsection, the kinematics of the dSphs in the MW and M31, for large 
values of the self-interacting parameter $\Lambda$ , are analyzed.
In the $\Lambda\gg1$ limit, as mentioned in section~\ref{sec:Big lambda}, the 
SFDM density profile is described by Equation~\ref{eq:TFL}.
In order to build the DM halos the value of the central DM density $\rho_{0}$ and the 
size of the SFDM halo $R_{max}$ must be fixed.

To study how well the circular velocity, and mass profile of the three 
groups of dSphs (M31, MW and ALL) are fit with a single SFDM halo in $\Lambda>>1$ 
regime the maximum likelihood fitting routine is modified, so that $\rho_{0}$ and 
$R_{max}$ are the free parameters. Then similarly to Equation~\ref{eq:max_prob} one 
has

\begin{multline}
\label{eq:max_prob2}
L_{TF} (\{r_{h,i},V_{c,i},\delta_{V_{c},i}\}|\rho_{0},R_{max}) = \\
\prod_{i=0}^{N} \frac{1}{\sqrt{2\pi} \delta_{V_{c},i}} 
\times exp \Biggl[-\frac{(V_{c,TF}-V_{c,i})^{2}}{2\delta_{V_{c},i}^2}\Biggr] ,
\end{multline}
where $V_{c,TF}$ is the circular velocity in the TFL.

Using the CMA-ES optimization routine described in Appendix~\ref{appendix}, 
I find that for the M31 dwarfs, the best fit is achieved for a value of $m_{\phi}^4/\lambda=3180$, 
which corresponds to a $\rho_{0} \sim 8\times10^7$~M$_{\odot}$~kpc$^{-3}$, a $R_{max}=0.87$~kpc, 
and at circular velocity value of $18.24$~km/s (see Table~\ref{table:2}). The latter value is 
approximately $13$\% higher than the one obtained for the $\Lambda=0$ case ($16.1$~km/s). 
The core radius (halo DM mass $M$) in the TFL for M31 dwarfs, is $0.52$~kpc ($6\times10^7$~M$_{\odot}$), 
very similar to the $0.45$~kpc ($5\times10^7$~M$_{\odot}$) find for the $\Lambda=0$ case.
The results are summarized in Table~\ref{table:2}. 

For the MW dwarfs the central density is the highest of the three groups, 
$1.8\times10^8$~M$_{\odot}$~kpc$^{-3}$, whereas the maximum DM radius for the MW dSphs is the smallest 
of the three groups, $R_{max}=0.59$~kpc (which corresponds to a core radius of $0.36$~kpc). 
Such a core radius is very similar to the one obtained for the $\Lambda=0$ case, and thus, it is 
large enough to explain the stellar substructures found within the UMi and the Sextans dSphs. 
The maximum circular velocity is $18.6$~km/s, again  approximately $11$\% 
higher than that found in the $\Lambda=0$ case ($16.7$~km/s). For the MW dwarfs a 
value of $m_{\phi}^4/\lambda=6788$ is obtained (i. e. $m_{\phi}/\lambda^{1/4}\approx9$). 

The upper panels of Figure~\ref{fig:FIG4} show the circular velocity as a function of the half-light 
radius, and DM halo mass as a function of the half-light radius (lower panels) for the 
SFDM model for $\Lambda=0$ and $\Lambda>>1$. 
The pink lines show the best fit for the SFDM model in the TFL ($\Lambda>>1$)
for each of the dSph data sets: ALL (left panels), M31 (central panels) and MW (right panels),
respectively. The shaded regions show the $1\sigma$ deviation. 
The results obtained for the $\Lambda=0$ case (light-blue), are also plotted in 
order to facilitate comparison between both SFDM profiles.

In the right panel of Figure~\ref{fig:FIG3} I show the seven dSphs which are in better agreement 
with a SFDM model in the TFL (such galaxies are plotted with pink symbols) also including
the previous studied DM profiles. The galaxies that are better reproduced with the SFDM in the
TFL are also mostly contained in the upper-right shaded gray region. This means that the
SFDM model in the TFL better fit galaxies with low mass-to-light ratios.

\subsection{The universal SFDM profile compared}
\label{sec:The universal SFDM profile compared}

In \cite{lora:12} we performed $N$-body simulations of the UMi dSph, which contains a 
cold-old stellar substructure \citep{kleyna:03}. We explored how the dissolution time-scale of such 
stellar substructure depends on the mass of the boson $m_{\phi}$. The boson mass range obtained 
from UMi's dynamics is $0.3\times10^{-22}<m_{\phi}<10^{-22}$~eV. 

Similarly, in \cite{lora:14} we investigated the dSph Sextans, which has two different old-cold stellar
substructures \citep{battaglia11,walker06}. For Sextans we require a mass for the SFDM boson of 
$0.12\times10^{-22}<m_{\phi}<8\times10^{-22}$~eV, in order to guarantee the survival of both stellar
substructures.

In Figure~\ref{fig:FIG1}, the corresponding mass of the boson is plotted (purple blue stars) for UMi 
and for Sextans. The error-bars denote the $1\sigma$ deviation. The best fit of the MW(triangle), 
M31(circle), and \scriptsize{ALL} \normalsize the sample (square) are also plotted.

For the Sextans dSph, all the values for the mass $M$ and the core radius $r_c$ in the light-gray
area in Figure~\ref{fig:FIG1} are permitted. In particular, those corresponding to the three best 
SFDM fits (MW, M31 and \scriptsize{ALL}\normalsize). The latter means that the dynamics of the 
Sextans dwarf, is in agreement with a unique SFDM mass profile. The UMi dSph is located above the 
MW, and \scriptsize{ALL} \normalsize fits in Figure~\ref{fig:FIG1}, being the most MW dSph 
restrictive case. However, it is still in agreement with the MW (circle) case at $1\sigma$ confidence.

My findings in this work in the TFL can be directly compared with the recent
results of \cite{diez-tejedor:14}. They constrained the parameters of the SFDM self-interacting 
parameter with the kinematics of the eight brightest dSph of the MW. 
They reported a preferred
value of $R_{max}\sim1$~kpc, which corresponds to $m_{\phi}/\lambda^{1/4}\sim7$~eV. In this work I
obtain for the MW sample a value of $R_{max}\sim0.6$~kpc, which corresponds to 
$m_{\phi}/\lambda^{1/4}\sim 9$~eV. It has to be noted that \cite{diez-tejedor:14} analyzed the eight
brightest dSph in the MW; in this work the MW sample was comprised of 19 dwarf galaxies. 
If the MW sample is restricted to the eight brightest dSph, I obtain a value  $R_{max}=0.57$~kpc, 
which corresponds to $m_{\phi}/\lambda^{1/4}=9.24$~eV. This result is still in agreement with the 
results obtained from the 19 MW dwarf sample, but  $m_{\phi}/\lambda^{1/4}$ (and $R_{max}$) is  
somewhat higher (lower) that the $m_{\phi}/\lambda^{1/4}\sim7$~eV ($R_{max}\sim1$~kpc) values 
reported by \cite{diez-tejedor:14}.

\textbf{It has to be noted that \cite{diez-tejedor:14} analyze the velocity dispersion of the 
eight brightest dSph galaxies in the MW, arguing that the stellar component of each of the 
eight galaxies are in dynamical equilibrium, and that the stellar distribution traces the DM 
distribution.}
\textbf{They fit three free parameters in their models: $R_{max}$, $M_{max}$ and
the orbital anisotropy of the stellar component of each galaxy, which is a difference
in the approach of this work. It is encouraging that, even with these differences, when the uncertainty 
averaged over the eight dSphs is taken into account in $Rmax$ (and thus $m_{\phi}/\lambda^{1/4}$) 
one obtains $R_{max}=1.05^{+0.3}_{-0.22}$ ($m_{\phi}/\lambda^{1/4}=6.82^{+0.87}_{-0.81}$), which 
is roughly in agreement with the findings reported in this work.}

\begin{figure}
\epsscale{1.00}
\plotone{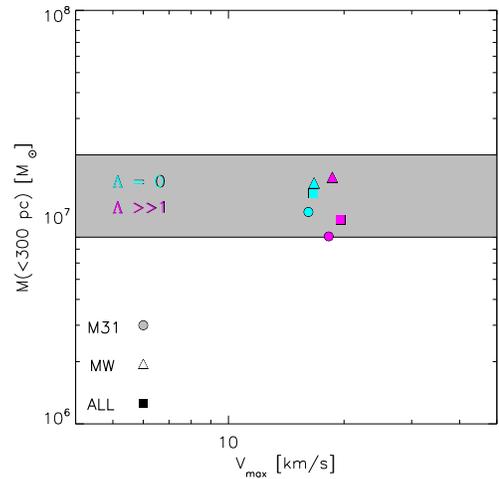}
\caption{The gray shaded region shows the results of \cite{strigari:08}. The resulting
mass within $300$~pc for $M31$ (circle), the $MW$ (triangle), and all the sample (square) 
are shown, for the $\Lambda=0$ case (light blue symbols) and the $\Lambda\gg1$ case (pink 
symbols).}
\label{fig:FIG5}
\end{figure}

\cite{strigari:08} suggested that dwarf galaxies satellites of the MW, have a common mass 
within $300$~pc of $\sim10^7$~M$_{\odot}$. They suggest that such a finding could shed light
in a characteristic scale for the clustering of dark matter.
In Figure~\ref{fig:FIG5} the gray shaded region shows \cite{strigari:08}'s results. Overplotted
are the results of this work. The light blue symbols correspond to the $\Lambda=0$ case, and 
the pink ones correspond to the $\lambda\gg1$ case. It is encouraging to find that for the best 
fits of the SFDM model for the M31 and MW dSph galaxies, the mass within $300$~pc is in agreement
with \cite{strigari:08} findings.

It has to be noted that the central mass density of the MW's dSphs range from $\rho_{0}\approx 0.03$ 
to $0.3$~M$_{\odot}$~pc$^{-3}$ \citep{kormendy:14,burkert:14}. It is encouraging that the latter values 
of the central density are not only in agreement with the central density obtained for the MW's sample 
($\sim0.18$~M$_{\odot}$~kpc$^{-3}$), but also with the whole sample 
($\sim0.1$~M$_{\odot}$~kpc$^{-3}$).

\section{Conclusions}
\label{sec:conclusions}

The very high dark-to-stellar mass ratios of dSph galaxies suggest that they are the most 
DM dominated objects in the universe, and therefore ideal laboratories to test any DM
alternative model. In this work I compare the kinematics of $22$ dSph galaxies satellites of M31, 
and $19$ dSph galaxies of the MW. I study also a third group of galaxies containing the sum of
all M31 and MW dSphs (ALL). 

I study the hypothesis that all dSph are embedded in dark matter halos with a same 
mass  \citep{walker:09,collins:14}, by fitting the dSph's half-light radius,
and the velocity at the half-light radius, to the SFDM model, in two different regimes:
when $\Lambda=0$ and when $\Lambda\gg1$. 

There is a very good agreement between the velocity at the half-light radius for the best fit 
for the ALL sample ($16.4$~km/s) from the $\Lambda=0$ case, with those obtained using an NFW ($16.2$~km/s),
and a cored ($15.6$~km/s) DM profile (see Equation~\ref{eq:core-nfw}). This corresponds to a mass
of the SFDM boson of $\sim4\times10^{-22}$~eV, which is in agreement with our previous findings
\citep{lora:12,lora:14}.

A higher value for the velocity at the half-light radius was found for the $\Lambda>>1$ case 
($19.6$~km/s) resulting in a  $m_{\phi}/\lambda^{1/4}\sim7.6$~eV. 

The M31-dSph galaxies with high luminosities and mass-to-light ratios ranging from $27$ to
$78$, are better reproduced with the SFDM model than with an NFW or core DM model (see light-blue 
symbols in Figure~\ref{fig:FIG3}). Hence, MW-dSph with very high mass-to-light ratios 
(ranging from $1500$ to $\sim100$) and very low values of the luminosity are in better agreement 
with a NFW/core DM profile than with a SFDM (see dark-blue and purple symbols in Figure~\ref{fig:FIG3}).

The mass within $300$~pc for the SFDM model (for both $\Lambda=0$ and $\gg1$) is in a very good 
agreement with \cite{strigari:08}'s findings. The latter suggests a universal SFDM halo 
mass for the dSph galaxies in the MW and M31. These results are encouraging, since also a unique
mass of the SFDM boson (for the $\Lambda=0$ case) of $\sim4.8\times10^{-22}$~eV is obtained. One
would expect a unique mass of the SFDM boson, and not to have a different SFDM boson mass for 
different dSph galaxies.


\acknowledgments
I would like to thank Michelle L. M. Collins and Nicolas F. Martin, for making their 
data available to me. 
I also thank Steffen Brinkmann and Colin W. Glass from HLRS, for making their
CMA routine available to me.
I thank Andreas Just, Juan Maga\~na and Avon Huxor for very helpful comments and discussions, that gave 
as a result an improved version of this paper. 
I gratefully acknowledges support from the Heidelberg University Innovation Fund FRONTIER.
%


\appendix
\section{The Evolution Strategy with Covariance Matrix Adaptation optimization}
\label{appendix}
The Evolution Strategy with Covariance Matrix Adaptation (CMA-ES) is a stochastic, derivative-free 
method for numerical optimization \citep{hansen:01,hansen:03}. It is robust 
and superior to derivative based search methods in rugged search landscapes, i.e.
containing noise, outliers, large derivatives etc. \\

An evolutionary strategy algorithm uses \textit{mutation} and \textit{selection} on a given
sample of solutions until a termination criterium is met. More precisely,
the algorithm is initialised with a population of $\mu$ candidate solutions in the solution space,
called parents. From these parents, $\lambda \geq \mu$ new candidate solutions (children) are created by mutation, i.e.
by scattering their properties (=genes) around the parents properties:

\begin{equation}
\mathbf{x}_i \sim \sigma  \mathcal{N}_i(\mathbf{m},\mathcal{C})\quad \mathrm{for} \ i = 1,\ldots,\lambda \mbox{ .}
\end{equation}
In the latter Equation $\mathbf{x}_i$ is the property vector of child $i$, $\mathbf{m}$ is the weighted mean 
property vector of the parents, $\sigma$ is the step size, $\mathcal{N}$ is the normal 
distribution around $\mathbf{m}$, which is distorted by the covariance matrix $\mathcal{C}$.
The children are evaluated by a fitness function ($f(\mathbf{x}): \mathbb{R}^n\rightarrow \mathbb{R}$)
and only the fittest $\mu$ children become the parents of the next evolutionary step.\\

CMA-ES enhances the evolutionary approach by updating the step size and the covariance matrix 
independently after each evolutionary step. For that purpose, the algorithm takes
into account the $\mu$ fittest children according to a given fitness function, the distance and
direction of the new weighted mean relative to the old one, and lastly, how many 
children achieved a better fitness than the best parent.\\

\begin{figure*}[t]
\epsscale{0.99}
\plotone{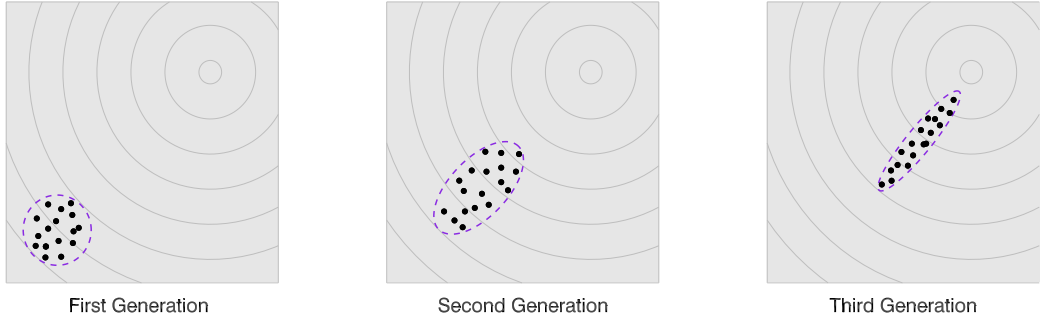}
\caption{Representation of the CMA optimization method.}
\label{fig:FIGapp}
\end{figure*}

The mean of the distribution is updated such that the likelihood of previously 
successful candidate solutions is maximized. The covariance matrix of the 
distribution is incrementally updated such that the likelihood of previously 
successful search steps (mutations) is increased. Both updates can be interpreted as a natural 
gradient descent. See Figure~\ref{fig:FIGapp}.\\

The adaptive algorithm design makes CMA-ES a black-box optimization strategy in the sense that 
no information about the solution landscape is required. The algorithm is furthermore robust
against transformations of the solution space, e.g. by exchanging the order of parameters in the
fitness function or different dimensional scaling of the solution space. Dependences between 
the solution candidate properties influence the covariance matrix, thus accelerating the search
for the optimal solution.

\end{document}